\documentclass[letterpaper,12pt]{article}

\begin{document}
\textbf{ \LARGE{Poincar\'e and Special Relativity}}
\vspace{8mm}

Emily Adlam

\textit{Department of Philosophy, The University of Oxford}

\vspace{6mm}
\textbf{ \Large{Abstract} }
\vspace{6mm}

Henri Poincar\'e's work on mathematical features of the Lorentz transformations was an important precursor to the development of special relativity. In this paper I compare the approaches taken by Poincar\'e and Einstein, aiming to come to an understanding of the philosophical ideas underlying their methods. In section (1) I assess Poincar\'e's contribution, concluding that although he inspired much of the mathematical formalism of special relativity, he cannot be credited with an overall conceptual grasp of the theory. In section (2) I investigate the origins of the two approaches, tracing differences to a disagreement about the appropriate direction for explanation in physics; I also discuss implications for modern controversies regarding explanation in the philosophy of special relativity. Finally, in section (3) I consider the links between Poincar\'e's philosophy and his science, arguing that apparent inconsistencies in his attitude to special relativity can be traced back to his acceptance of a `convenience thesis' regarding conventions.

\vspace{6mm}
\textbf{ \Large{Keywords} }
\vspace{3mm}

Poincare; Einstein; special relativity; explanation; conventionalism; Lorentz transformations

\vspace{45mm}
\textbf{ \Large{1 Did Poincar\'e Discover Special Relativity?} }
\vspace{6mm}

Poincar\'e's work introduced many ideas that subsequently became important in special relativity, and on a cursory inspection it may seem that his 1905 and 1906 papers (written before Einstein's landmark paper was published) already contain most of the major features of the theory: he had the correct equations for the Lorentz transformations, articulated the relativity principle, derived the correct relativistic transformations for force and charge density, and found the rule for relativistic composition of velocities. His conceptual approach also seems to have much in common with Einstein; for instance, in 1898 he gave a derivation of Lorentz's `local time' coordinate that closely mirrors Einstein's derivation of the Lorentz transformations, particularly in its use of an identical convention for synchronising spatially separated clocks. Nonetheless, a closer reading of Poincar\'e's papers reveals that his understanding of both the relativity principle and the Lorentz transformations is significantly more limited than Einstein's.  

\vspace{3mm}
\textbf{ 1.1 The Relativity Principle}
\vspace{3mm}

One element which links the work of both Poincar\'e and Einstein is a preoccupation with the principle of relativity. But it is important to be aware that Einstein and Poincar\'e were not working with precisely the same principle. Compare their two fomulations: 

\vspace{2mm}

Poincar\'e : `the laws of physical phenomena must be the same for a motionless observer and for an observer experiencing uniform motion along a straight line.' (1904) \footnote{In his 1902 essay 'Relative and Absolute Motion', Poincar\'e gave a different formulation of the principle of relativity, omitting the reference to an observer: `the movement of any system whatever ought to obey the same laws, whether it is referred to fixed axes or to the movable axes with are implied in uniform motion in a straight line.' (1902, p.111) But this version appears in a philosophical paper rather than a scientific one, and as we shall see, Poincar\'e's scientific views must be kept separate from his philosophical ones.}  

\vspace{2mm}

Einstein: `The laws by which the states of physical systems undergo change are not affected, whether these changes of state be referred to the one or the other of two systems of co-ordinates in uniform translatory motion.' (1905)

\vspace{2mm}

The crucial difference between these formulations is that Poincar\'e finds it necessary to refer to an observer, while Einstein does not. The difference has been noted by  Katzir: `In contrast to Einstein, who denied the existence of absolute motion, Poincar\'e denied the possibility to detect it.'  (2005) As a result Einstein's principle leads to stronger constraints: for Einstein, there can be no difference at all between the forms of the laws of nature in different inertial frames, whereas Poincar\'e can accept that the laws of nature take one form relative to a privileged frame and a more complicated form relative to all other frames, provided they work together in such a way that this difference between frames does not have any observable consequences. However, the importance to be attached to this distinction turns upon our understanding of what is meant by a 'law of nature.' Poincar\'e was famously a conventionalist, and therefore held a view of lawhood which may appear to make the two principles equivalent after all. He claimed that the modern notion of a law is `a constant relation between the phenomena of today and tomorrow, i.e. a differential equation,' (1904) which suggests that in his view, a law is nothing over and above relations between observable phenomena. If taken seriously, this suggests that the true form of the laws of nature cannot be different to the phenomenlogical laws that would be formulated on the basis of observations within any given frame: there is no distinction to be made between `the laws for an observer' and the actual laws of nature. 

But if Poincar\'e were consistent about this view of lawhood, his acceptance of the principle of relativity would surely force him to abandon the notion of a privileged frame of reference: if there are no laws of nature above and beyond relations between what is observable, and if observation can never disclose a privileged frame of reference to us, it follows that there is no such frame. Yet Poincar\'e's theories remain tied to the notion of a a privileged frame of reference; for instance, in his 1906 paper `On the Dynamics of an Electron,' he gives an analysis of the motion of an electron in which he continues to refer to the `real electron,' meaning the electron as it appears to observers in the ether rest frame. Whatever Poincar\'e might believe in the context of philosophy, in the context of his scientific work he assumes that at least some of the actual laws of nature are distinct from the laws formulated by observers. In this case, the actual laws of nature, which single out a privileged rest frame, conspire to produce the same observable effects in all inertial reference frames so that all observers in such reference frames will formulate the same laws on the basis of their observations. Further evidence that this was Poincar\'e's view can be found in the status he accords to the relativity principle, which he takes to be an empirical result, an inductive generalisation from the null results of ether drag experiments such as the Fizeau experiment, the Trouton-Noble experiment, and the Michelson-Morley experiment. Thus, for example, he writes the principle is`so far in agreement with experiment ... (but may) be later confirmed or disproved by more accurate tests.' (Logunov 2001, p. 15). Induction from these experimental results justifies the claim that all inertial frames of reference appear identical to observers, but not necessarily the stronger claim that all such frames of reference are in fact identical. We are therefore justified in making a distinction between `the laws for an observer' which are the object of Poincar\'e's relativity principle, and the laws of nature simpliciter which are the object of Einstein's. 

Nor is this distinction a trivial one. An important motivation for Zahar's claim (2001) that Poincar\'e should be credited as the discoverer of special relativity is that he `obtained, as a first heuristic component of his programme, the \textit{Principle of Lorentz-covariance} which is in effect a \textit{symmetry} requirement,' and this heuristic has been crucial to the development of special relativity. However, because Poincar\'e's relativity principle is subtly different to Einstein's, he also had a different understanding of the principle of Lorentz covariance. Like Einstein he imposes the requirement of Lorentz covariance on all the fundamental equations of nature, but the equations in question are still understood relative to the ether rest frame and are never referred to any other frame of reference, and therefore for Poincar\'e, Lorentz covariance is a condition on the solutions to some set of equations relative to a single reference frame. For Einstein on the other hand, the motivation behind Lorentz covariance is that the equations should be unaffected by the coordinate transformations because the principle of relativity demands that they should actually take the same form in all frames, because there is no special reference frame with respect to which the formulation of the laws takes conceptual priority. Thus although Poincar\'e's Lorentz covariance condition is mathematically equivalent to Einstein's, it is very differently motivated and therefore as a heuristic it offers a different sort of guidance. It is therefore not entirely accurate to credit Poincar\'e with the invention of the methodology of later research in special relativity, because although he provided the mathematical background for this heuristic, he did not fully appreciate its physical interpretation.

\vspace{3mm}
\textbf{1.2 The Lorentz Transformations} 
\vspace{3mm}

Since the Lorentz transformations are at the heart of the theory of special relativity, Poincar\'e cannot fairly be said to have anticipated Einstein unless his understanding of the Lorentz transformation was sufficiently close to Einstein's. It is often assumed that Poincar\'e, like Einstein, thought of the Lorentz transformations as a a procedure for changing between different coordinate systems - so for instance, Brown claims that 'Poincar\'e was the first to use the generalized relativity principle as a constraint on the form of the coordinate transformations,' (2005) and Zahar writes that `Poincar\'e eliminated (the Galilean coordinates) in favour of the Lorentz-transformation, a transformation which goes directly from the rest frame to the effective coordinates.' (1989, p. 174) However, I submit that Poincar\'e never regarded the Lorentz transformations as coordinate transformations: he saw them as a calculational device rather than a physical relationship between actual frames of reference. 

A preliminary reason to be sceptical about the depth of Poincar\'e's understanding of the Lorentz transformations is that neither he nor Lorentz attempted to give a derivation for them; the equations were simply selected by Lorentz on the grounds that they happen to preserve the form of the Maxwell equations. It is doubtful that anyone could fully appreciate the way in which the Lorentz transformations codify the behaviour of moving rods and clocks without seeing how the equations can be obtained directly from an operationalist approach to changing reference frames, as Einstein demonstrates in his 1905 derivation. Moreover, the reasoning Poincar\'e uses when working with the transformations is consistently abstract and mathematical rather than physical. In his 1906 paper `On the Dynamics of the Electron' he offers a derivation of the the factor $L$, which is common to all the transformed coordinates and is left undetermined by the requirement that the transformations should preserve the form of the Maxwell equations. But his derivation involves no physical considerations: he reaches his conclusion simply by applying certain mathematical constraints. For all the elegance of this method, it seems an unusual approach to take in attempting to prove something which, if the Lorentz transformations are interpreted physically, determines the extent to which a moving object contracts in the transverse direction, and is therefore an important empirical feature of the world. The strategy would not be unjustifiable, since there are intuitively plausible symmetry principles which offer physical reasons to believe that the Lorentz transformations should form a group, but Poincar\'e never invokes them; he merely asserts that `We must regard $L$ as being a function of $\beta$, the function being chosen so that this partial group, which will be denoted by $P$, is itself a group.' (Logunov 2001, p. 41). This approach seems more in harmony with the idea that the Lorentz transformations are a convenient calculational tool, since then we are free to stipulate that they should have any mathematical property which we find convenient and which does not interfere with their role in calculation. Moreover, it is also noticeable that neither Poincar\'e nor any of his peers ever considered possible alternatives to the Lorentz transformations. After all, Lorentz's approach simply involved finding a set of transformations which retain the form of the Maxwell equations -  he gave no reason to think the set is unique in this regard. It seems at least prima facie plausible that there might be other transformations with the same property (indeed, we know that the Lorentz transformations are only a subset of the covariance group of the Maxwell equations), and therefore if the transformations were being interpreted realistically, it would be natural to ask whether we have found the transformations which real systems actually obey. But if the transformations are merely a convenient calculational device, this question has no meaning; therefore we can understand why the question was never raised if we accept that Poincar\'e and his peers did not see beyond the mathematical interpretation. 

Thus it seems that Poincar\'e's interpretation of the Lorentz transformations was rather far removed from Einstein's physical understanding of them. Nonetheless, if the ways in which he used the transformations are sufficiently similar to their applications in special relativity, perhaps he can be said to have had a partial understanding of the transformations in virtue of his appreciation of their practical role. It is therefore important to examine Poincar\'e's ideas about the function of the Lorentz transformations, as distinct from their theoretical origins. Clearly the transformations express a relationship between two sets of coordinates: but what do these coordinates signify? For Einstein, the coordinates $x$, $y$, $z$, $t$ describe the spatiotemporal location of some event with respect to an inertial frame $S$, and the transformed coordinates $x'$, $y'$, $z'$, $t'$ describe the spatiotemporal location of the same event as it would be measured by an observer at rest in a frame moving at speed v with respect to $S$ (provided that the Einstein synchrony convention is used). But for Lorentz and Poincar\'e, the transformation is applied very differently. We consider a physical system which is in motion with respect to the ether rest frame and suppose that the coordinates $x$, $y$, $z$, $t$ describe configurations of the parts of that system with respect to the ether frame. When we apply the Lorentz transformations we obtain transformed coordinates $x'$, $y'$, $z'$ and $t'$ which describe what Poincar\'e calls the `ideal' system, in contrast with the `real' system. For example, in a discussion of various models of the electron in his 1906 paper, Poincar\'e considers a hypothesis due to Abraham which asserts that electron is spherical and nondeformable -  that is, it does not undergo length contraction. He writes that `upon applying the Lorentz transformation, since the real (moving) electron is spherical, the ideal electron will become an ellipsoid.' (Logunov 2001, p. 49) Here the Lorentz transformations are applied to a real, moving, spherical electron to give an alternative coordinatization which makes it an ellipsoid. Crucially, this alternative description is not referred to any physical reference frame, nor given any physical interpretation. Indeed, Poincar\'e never associates the transformations with the process of changing reference frames, nor does he demonstrate any awareness that there could be interesting physics related to changes in reference frames - his analyses are always carried out in the ether rest frame and the motion involved is always absolute. 

But if the Lorentz transformations do not express relationships between two physical reference frames, what is their purpose? In the theory advocated by Poincar\'e, their main function is to permit the formulation of the hypothesis that when a system is set in motion with respect to the ether, it undergoes certain changes in its configuration such that when we apply the Lorentz transformation, we obtain an `ideal' system which is at rest in a corresponding `ideal' coordinate system and has the same configuration as the real system has when it is at rest in the ether rest frame. Clearly the real system can be recovered by applying the inverse transformations, and since the Lorentz transformations form a group, it follows that the real moving system is related to the ideal system (i.e. the system at rest) by a Lorentz transformation - that is, the hypothesis amounts to the requirement that when a system is set in motion it turns into the corresponding Lorentz transformed system. Unfortunately neither Lorentz nor Poincar\'e ever gave a complete and explicit statement of this hypothesis, which Janssen calls the `generalised contraction hypothesis,' but it is undoubtedly necessary if the use of the Lorentz transformations is to achieve the avowed aim of preserving the relativity principle: as Janssen points out, `the configuration of a material system at rest in the ether will have to change upon setting the system in motion if it is to generate the electromagnetic field configuration in the moving frame that is the corresponding state of the electromagnetic field configuration generated by the system at rest in the ether.' (1995) We know that the system at rest obeys the Maxwell equations, and the Lorentz transformations were chosen specifically to preserve Maxwell's equations, so if the moving system changes in the way hypothesized, it will continue to obey the Maxwell equations. If it is the case that not only the Maxwell equations but all laws of nature are left unchanged by Lorentz transformations, we can prove that the relativity principle will never be violated, because the same laws of nature that are obeyed by the system at rest will also be obeyed by the moving system: as Poincar\'e puts it, `two systems, of which one is fixed and the other is in translatory motion, become exact images of each other.' (Logunov 2001, p. 16) Thus the purpose of the Lorentz transformations, in Poincar\'e's theory, is simply to enable us to determine what changes the system must undergo when it moves with respect to the ether in order that it will still satisfy the Maxwell equations and any other Lorentz covariant equations. 

It is important to be aware that although Poincar\'e used the Lorentz transformations to ascertain the changes necessary to preserve the relativity principle, he did not take the transformations to be a cause or an explanation of these changes - he thought it necessary to postulate distinct explanations for length contraction, local time, and other such phenomena. He dealt with the contraction dynamically, writing that `a special force must be invoked to account for both the contraction and the constancy of two of the axes,' (Logunov 2001, p. 18). Accounting for the phenomenon of local time is less straightforward: Lorentz avoided the issue by asserting that the difference between local time and actual time is too small to be significant, but Poincar\'e realised that without some way of accounting for the required change in the temporal coordinate, the theory would conform to the relativity principle only approximately, which he found unacceptable. He therefore sought to show that if clocks were synchronised according to a particular synchrony convention, equivalent to the one which Einstein later adopted, clocks in moving systems would be synchronised according to local time rather than real time, thus effecting the required change. \footnote{However, Poincar\'e's derivation only gives the equation for local time which appeared in the original, first-order version of the transformations, not the exact transformations which Lorentz published in his paper 'Simplified Theory of Electrical and Optical Phenomena in Moving Systems' (1899): Poincar\'e's local time is given by the formula $ t' = t - \frac{vx}{c^2}$, whereas the exact time transformation is given by $t" = \gamma(t - \frac{vx}{c^2})$. The derivation was therefore adequate in 1898 when Poincar\'e first provided it, but had gone out of date by the time of his 1906 paper. However, no steps were taken by either Lorentz or Poincar\'e to show that the time coordinate would change in a way consistent with the exact version of the transformations; Poincar\'e must surely have been dissatisfied with this result, but perhaps he hoped to resort to Lorentz's earlier strategy and claim that the new temporal coordinate required by the exact transformation would be so close to the local time that the difference would not matter – an assumption that would be justified in most cases, since $\gamma$ is close to 1 unless the velocities involved are very large.}

In summary, Poincar\'e's use of the Lorentz transformations differed from Einstein's in two key ways. First, for Poincar\'e, the transformations are not used to give relations between any two inertial frames of reference - they are defined only relative to the ether rest frame, so that the velocity v appearing in the formula must always refer to a velocity with respect to the ether rest frame, i.e. an absolute velocity. Moreover, even if we restrict ourselves to transformations involving the ether rest frame, Poincar\'e's usage does not coincide with Einstein's, since for Einstein the transformations express a relationship between coordinate systems, whereas for Poincar\'e they are merely a means of predicting the physical changes that a system undergoes when set in motion relative to the ether. Thus Poincar\'e not only failed to give a physical interpretation to the Lorentz transformations, he also failed to appreciate the full range of situations in which they can be applied.

\vspace{3mm}
\textbf{1.3 Assessment}
\vspace{3mm}

In light of Poincar\'e's limited understanding of the relativity principle and the Lorentz transformations, it seems inaccurate to say that he had any significant intimations of special relativity before Einstein's 1905 papers. This is not to deny that he made extremely important contributions to the development of the theory, but his achievements in this area were largely mathematical: formulating the notion of the Lorentz group and finding its invariants, formulating the notion of a four-vector and finding quantities that transform like four-vectors, interpreting the Lorentz transformations as rotations in four-dimensional space. These are results that follow from the mathematical structure of the equations, not from any physical understanding of their significance; they paved the way for the powerful mathematical formalism developed by later workers in the field, but did not provide the essential physical insight that provides the formalism with its application. Special relativity is first and foremost a physical theory, and in the absence of an understanding of the physical significance of the Lorentz transformations, Poincar\'e cannot be said to have formulated a theory approximating special relativity. 

\vspace{6mm}
\textbf{ \Large{2 Explanation in Special Relativity} }

\vspace{6mm}
\textbf{2.1 The direction of explanation} 
\vspace{3mm}

There is some controversy over the true nature of the disagreement between the theories of Poincar\'e and Einstein. The difference is not empirical - Janssen (1995) shows that although the Lorentz-Poincar\'e theory before 1905 is not exactly empirically equivalent to special relativity, it can be made so with minimal alterations. It might seem that the main point of difference is ontological: Poincar\'e's theory makes essential reference to the ether and thus to a privileged rest frame, while Einstein's theory does not. But we should not attach too much import to this fact, for Einstein was careful to point out that his theory does not actually rule out the existence of the ether. Another popular view, supported by Goldberg (1967), Miller (1973) and Hirosige (1976), is that the differences stem from Poincar\'e's commitment to the electromagnetic world-picture, upon which the only basic constituents of the world are charged particles and electromagnetic fields. But this does not seem to account completely for the distinctions between the theories; after all, the principle of relativity and the light postulate could certainly be true even in a wholly electromagnetic world and Lorentz covariance could still be derived from them, so such a commitment would not suffice to prevent Poincar\'e from taking Einstein's approach. Moreover, Katzir (2005) points out that in his 1906 paper, Poincar\'e accepts Lorentz's model of the electron, which is compatible with the relativity principle but not the electromagnetic worldview, over Abraham's model, which is compatible with the electromagnetic world view but not the relativity principle, and goes on to invoke the relativity principle as the reason for his choice. This demonstrates that Poincar\'e was willing to put the relativity principle above any attachment he may have had to the electromagnetic worldview, and therefore that view should not have interfered with his work on the relativity principle. 

Nonetheless I think the intuition that the electromagnetic worldview plays some role here does contain an element of truth. I suggest that the differentiating factor was not Poincar\'e's commitment to the  electromagnetic world-picture itself, but to the explanatory strategy associated with it. Although he was willing to accept the existence of particles which are not charged and forces which are not electromagnetic in nature, he remained faithful to the underlying motivation for the electromagnetic picture, which is that all observable phenomena should be be accounted for by appeal to the nature of the fundamental particles and forces. This idea was certainly not unique to proponents of the electromagnetic worldview; a similar motivation lies behind the mechanical world-picture, upon which all macroscopic phenomena are produced by Newtonian interactions between moving microscopic particles. Indeed, a commitment to the explanation of the macroscopic in terms of the microscopic seems to have been a general characteristic of physics in the era leading up to Einstein - for example, beginning with Boltzmann's derivation of the relationship between entropy and multiplicity in 1875, it was an ongoing project to show that the macroscopic laws of thermodynamics could be derived from assumptions about the microscopic world by the methods of statistical mechanics. 

This explanatory strategy is a consistent feature of Poincar\'e's work: witness his assumption that the physical changes predicted by using the Lorentz transformation must be explained by piecewise derivation from force laws and microscopic phenomena. It is therefore not surprising that Poincar\'e never thought to view the relativity principle as explanatory in and of itself; as Katzir puts it: `instead of deducing consequences from (the relativity principle), he used it mainly to confirm or refute various hypotheses.' Poincar\'e regarded the principle rather like a general summary of the empirical evidence, such that that theories which violated it could be taken to have been indirectly disconfirmed. For instance, in his 1905 paper, he which he offers a proof that `Lorentz's hypothesis (about the contraction of the moving electron) is the only one which is compatible with the impossibility of manifesting absolute motion,' and claims that `Lorentz's analysis is thus fully confirmed.' (Logunov 2001, p. 62). The principle of relativity thus functions as supporting evidence for the contraction hypothesis, but not as an explanation for the contraction, since Poincar\'e goes on to offer an entirely separate explanation in terms of a force law, insisting that if one believes the electron contracts, `one must admit ... the existence of a supplementary potential proportional to the volume of the electron.' Moreover, even after the publication of Einstein's paper Poincar\'e did not accept the use of the relativity principle to explain such phenomena: Pais (1982) emphasizes the fact that even in 1908, Poincar\'e was unwilling to take length contraction as a consequence of the relativity principle together with the light postulate. According to Poincar\'e and his peers, all legitimate scientific explanations ought to involve an appeal to microscopic laws: in the context of their notion of explanation, the relativity principle was simply not the right kind of thing to function as an explanans. 

Einstein's 1905 paper was revolutionary precisely because he broke with the long-standing tradition of explaining the macroscopic in terms of the microscopic. Rather than taking force laws as fundamental, he made the relativity principle the basic axiom of his theory and used it to derive constraints on the form of the laws governing phenomena at both the microscopic and macroscopic levels.  It must be emphasized that this difference in approach is not merely a trivial disagreement about how the theory is best presented, for the approach adopted by Einstein not only neglects to derive the relativity principle from more fundamental hypotheses, but actually rules out the possibility of doing so. If we accept the strategy of using the relativity principle to prove that all the fundamental equations of nature must be Lorentz covariant, it cannot itself be understood as a consequence of the Lorentz covariance of the equations of any present or future theory. Therefore inside this explanatory framework, we must either simply take the relativity principle as given without questioning why it is true, or justify it in terms of some fact about the world which is not itself a consequence of the principle. The only plausible candidates seem to be facts about space: either the assertion that there is no substantive space (or spacetime), or the less extreme claim that whatever space might be, it is not the kind of thing which could pick out a preferred reference frame and thereby render the relativity principle false. But it is important to be clear that choosing this form of explanation does not commit us to any ontological position on the nature of space, for the relativity principle is fundamentally a negative existential claim to the effect that no preferred reference frame exists, and in most cases we do not think it necessary to provide any justification for existential claims, particularly negative ones - after all, it would seem bizarre to ask why it is the case that unicorns do not exist. Moreoever, the very notion of explanation requries that there are certain facts about the world which, at least in the correct context, do not stand in need of explanation and can therefore be used to explain other facts. By accepting the explanatory strategy offered by Einstein we are making a free choice to adopt the principle of relativity as fundamental, circumventing any demand for any further justification. 

This means that the scientific community's adoption of special relativity involved not only a change in theory, but a change in accepted notions of what is to be expected and what requires explanation. For an ether theorist, or anyone committed to the existence of a privileged rest frame, both absolute velocity and absolute acceleration are real features of the world and it is therefore prima facie to be expected that we should be able to detect them. The fact that we cannot detect absolute velocity then seems to stand in need of explanation, hence the intuition on the part of Poincar\'e and his peers that it is important to present a derivation of the relativity principle. But suppose we discard the notion of an absolute rest frame; it follows that there is no such thing as absolute velocity. Our inability to detect absolute velocity is therefore entirely to be expected, and the relativity principle requires no derivation. \footnote{Notice that we can rid ourselves of absolute velocity without also discarding absolute acceleration, since we can claim that all inertial reference frames are equivalent without getting rid of the notion of inertial reference frames altogether. If we choose to justify the claim that there is no privileged rest frame by the relationalist strategy of denying the existence or causal efficacy of absolute spatial structure, then the fact that we can apparently detect absolute acceleration will stand in need of explanation, which might for instance be offered by taking acceleration to be relative to the total distribution of mass in the universe (see Barbour 2001), but the explanatory strategy offered here is perfectly coherent without any such extension.}

Such changes in our expectations are, I suggest, often important aspects of progression in physics. Prior to the formulation of the law of inertia by Galileo and its popularisation by Newton, the physics of motion was based on Aristotle's idea that `everything that is in motion must be moved by something.' (Physics, Book VII) which made it natural to take the view that whenever an object persists in a state of uniform motion, this phenomenon in need of explanation by some force. But the law of inertia produced a significant change in our ideas about what is to be expected and what requires explanation in physics: the law implies that uniform motion requires no explanation, while accelerations, including the process of slowing down to rest, do require explanation by some force. Thus there is a precedent for the kind of explanatory revolution that Einstein introduced; changes in explanatory strategy can be expected to yield interesting changes in physical theory. 

\vspace{3mm}
\textbf{2.2 The derivation of $L$ = 1}
\vspace{3mm}

I shall now consider a concrete example of the different approaches taken by Einstein and Poincar\'e.  For the purposes of clarity, it it best to choose an example where the mathematical procedures used are similar, in order to better show up the conceptual differences. Therefore we will be looking at the proofs given by Einstein and Poincar\'e, in their 1905 and 1906 papers respectively, for the fact that the multiplicative factor $L$ in the Lorentz transformations is equal to one. Mathematically, the two derivations are very similar, but the underlying ideas are quite distinct. 
 
Poincar\'e's derivation relies on the mathematical theory of groups. He first defines a set of transformations which are identical to the Lorentz transformations in units of c = 1, up to a scale factor $L$: 
(1)
\[ x' = \gamma L (x - \beta t) \]
\[ y' = Ly \]
\[ z' = Lz \]
\[ t' = \gamma L (t - \beta x) \]
No assumptions are made about any relationship between $\beta$ and $L$. Poincar\'e shows that two such transformations in successsion are equivalent to a third transformation of the same form: 
(2)
\[ x'' = \gamma '' L'' (x - \beta'' t) \]
\[ y'' = L''y \]
\[ z'' = L''z \]
\[ t'' = \gamma '' L'' (t - \beta'' x) \]
where, if $\gamma$, $\beta$ and $L$ are the parameters used in the first transformation and  $\gamma$', $\beta$' and $L'$ are the parameters used in the second, then 
 \[ \gamma '' = \gamma \gamma'(1 + \beta \beta ') \]
 \[ \beta'' = \frac{\beta + \beta'}{1 + \beta \beta'} \]
\[ L'' = L L'' \]

This proves that this unrestricted set of transformations obeys the closure requirement for groups. 
There are other conditions that must be obeyed if a set is to form a group - it must exhibit associativity and contain an identity element as well as the inverse of each of its member - and Poincar\'e does not explicitly prove that this set satisfies these conditions, but it is reasonably easy to see that they are in fact satisfied, and thus the set forms a group, which Poincar\'e calls the Lorentz group. 

He next considers a subset $P$ of these transformations such that for all members of $P$, $L$ is a function of $\beta$, and asks what this function must be in order that $P$ should also be a group. He observes that we must be able to rotate the system through 180 degrees, thereby changing the signs of $x$, $x'$, $z$ and $z'$, and the transformation relating these new coordinates should still be a member of the group. The equations for this transformation are given by: 
(3)
\[ x' = \gamma L (x + \beta t) \]
\[ y' = L y \]
\[ z' = L z \]
\[ t' = \gamma L (t + \beta x) \]
Compare this to the transformation equations obtained by using -$\beta$ instead of $\beta$: 
(4)
\[ x' = \gamma L_2(x + \beta t) \]
\[ y' = L_2 y \]
\[ z' = L_2 z \]
\[ t' = \gamma L_2 (t + \beta x) \]

Because we require that $L$ should be a function of $\beta$, there is a unique set of equations corresponding to each distinct value for $\beta$, which means that (3) and (4) must be identical, and therefore $L = L_2$, i.e. $L$ does not change when we reverse the sign of $\beta$. 

Now we return to transformations (1) and find their inverse:
(5) 
\[ x' = \frac{\gamma}{L}(x + \beta t) \]
\[ y' = \frac{y}{L} \]
\[ z' = \frac{z}{L} \]
\[ t' = \frac{\gamma}{L} (t + \beta x) \]

One of the conditions in the definition of a group is that the inverse of every element of the group is also an element of the group, and therefore transformations (5) must also belong to the group. We have defined this group such that $L$ is a function of $\beta$  only, and therefore the transformations (3) and (5) must be identical; it follows that $L = 1/L$, and therefore $L = \pm 1$. Poincar\'e does not discuss the possibility that $L$ might be negative, but Zahar (2001, p. 34) points out that taking $L = -1$ leads to a violation of the requirement that the group be closed under composition of transformations, since composing two transformations with $L = -1$ would lead to a third transformation with $L = 1$. There is an alternative justification: assuming that the value of $L$ must be either $1$ or $-1$, $L = 1$ is the only solution consistent with the transformations reverting to the identity transformation in the limit $v = 0$' (see for example Brown 2005).  These suggestions leave open the possibility that $L$ should be $1$ in some cases and $-1$ in others, but discontinuities in the value of $L$ seem intuitively objectionable, and since Einstein makes similar assumptions, such intuitions must be fairly widespread. 

The interesting point is why Poincar\'e thinks it is important for $P$ to form a group. In a reconstruction of the proof, Zahar writes `In view of the relativity principle, all allowable frames are equivalent, which entails that $P$ must form a group.' (1981, p. 191) This is the appropriate modern justification for the group requirement, but it is not clear that it is a correct description of Poincar\'e's thought process, since he never mentions the relativity principle in the course of his derivation. Of course, he might have thought its relevance was so obvious that the link did not need to be made explicit, but this seems unlikely in view of his understanding of the Lorentz transformations. The modern justification for the group requirement can be expanded as follows: if all frames are equivalent, the transformation from frame $A$ to frame $B$ should have the same form as the transformations from frame $B$ to $C$ and from $A$ to $C$, because otherwise moving objects will undergo different changes in constitution when they are in motion relative to different frames, which would contradict the principle of relativity, since it would provide us with a means of distinguishing between the two frames. Thus the two successive transformations from $A$ to $B$ and $B$ to $C$ should be equivalent to a single transformation from $A$ to $C$, and therefore two successive Lorentz transformations are equivalent to a third individual Lorentz transformation, which is equivalent to the closure requirement imposed on groups. Thus to recognise that the relativity principle implies that the Lorentz transformations must form a group, we must understand the Lorentz transformations as describing a relationship between the coordinate systems of any two inertial frames. But there is no evidence that Poincar\'e ever thought of the transformations in this way - his use of the transformations is limited to cases where a system is moving with some given velocity with respect to the ether rest frame, and even then, we have seen that he does not interpret the transformations as expressing a relationship between the coordinates of the two frames, but as a device for establishing how the system must change in order that it should continue to obey the Maxwell equations. The only occasion on which he ever considers two successive transformations is in the context of his mathematical proof that the unrestricted Lorentz transformations form a group, and since these are the unrestricted transformations which do not correspond to any physical transformations, their composition cannot be given any physical interpretation. Thus Poincar\'e shows no awareness that it would be meaningful to apply two successive Lorentz transformations; and if he was not aware of such a possibility, the relativity principle cannot be his justification for the assertion that $P$ should form a group. Therefore it seems more likely that he makes this stipulation for reasons of mathematical elegance and theoretical usefulness, since treating the Lorentz transformations as a group proves very productive later in the same paper when he makes various deductions about a possible Lorentz covaraint theory of gravity by considering the invariants of the group. 

Einstein's approach involves very similar mathematics: he too takes the inverse transformation and equates it to the transformation for –$\beta$, thereby proving that $L = 1/L$. But his reasons for asserting this equality are quite different to Poincar\'e's. Because Einstein has all along interpreted the transformations as a procedure for changing between the coordinate systems associated with observers in relative motion, it is reasonable for him to claim that if we apply the transformation for $\beta$ followed by the transformation for - $\beta$, we end up back in the same system in which we started, and therefore the transformation for  - $\beta$ must be the inverse of the transformation for $\beta$. It is true that this depends on an additional assumption of reciprocity - if system one has velocity $v$ with respect to system two, then system two has velocity - $v$ with respect to system one - but reciprocity follows immediately from the Einstein synchronisation convention which is used in the derivation, so this assumptions is justified. On the other hand, Poincar\'e arrives at this equality by finding the inverse transformations, noticing that they have the same form as transformations (3) except that $L$ is replaced by $1/L$, and applying the condition that $L$ is to be a function of $\beta$ only to reach the conclusion that $L$ must be the same in both cases, i.e. the transformations must be identical and therefore $L = 1/L$. No physical interpretation is given to transformations (3). Indeed, if it is true that Poincar\'e thought of the Lorentz transformations as giving a procedure for finding descriptions of systems in a certain state of motion with respect to the ether rest frame, a proof of the nature of Einstein's would have been outside his conceptual grasp - it would not even have occurred to him that one could apply the transformations and then apply them again in the new frame to get back to the original frame. 

Furthermore, it is noticeable that while Poincar\'e's first step is to give a proof in terms of mathematical structure that $L$ does not change when the sign of $\beta$ changes, Einstein merely assumes this `from reasons of symmetry.' (1905) It may seem that this approach bears out Lorentz's complaint that `Einstein simply postulates what we have deduced,' (1916) but in fact, the difference mirrors a profound divergence between the way that Einstein and Poincar\'e understand the derivation. For Poincar\'e, the transformations are primarily a mathematical device, and therefore a claim about the relation between $L$ and $\beta$ can only be justified mathematically. Einstein, on the other hand, was thinking about the transformations in a physical manner, and therefore physical symmetries are relevant, in particular the isotropy of space. Once the isotropy of space has been assumed and the transformations have been given a physical interpretation, a proof along the lines of Poincar\'e's is superfluous - Einstein notes that $L$ corresponds to the degree of transverse contraction, and then points out that if $L$ varied depending on the direction of motion, the degree of contraction would vary depending on the direction of motion, which would violate the isotropy of space. 

\vspace{3mm}
\textbf{ 2.3 Einstein's view}
\vspace{3mm}

I have claimed that the fundamental insight of Einstein's 1905 paper is that the relativity principle can be taken as an explanans rather than an explanandum. However, I do not mean to assert that Einstein ever saw the matter in this light; indeed, I am inclined to favour the view that he did not. As a result of his study of the photoelectric effect, culminating in the paper `On a Heuristic Viewpoint Concerning the Production and Transformation of Light,' which he published earlier in 1905, he was well aware that the accepted laws of electrodynamics could not be exactly true, and he later claimed that he consequently `despaired of the possibility of discovering the true laws by means of constructive efforts based on known facts.' (1949) Thus he put off the task of deriving the principle of relativity until a more detailed theory could be found, and simply assumed the truth of that principle in order to examine some of its consequences. It is clear from his later writings that he did not intend the relativity principle to remain a basic axiom, since in his Autobiographical Notes (1949) he characterises special relativity as a `principle theory' akin to thermodynamics, with the relativity principle being analogous to the second law of thermodynamics. We do not imagine that the laws of thermodynamics are primitive facts about the world - they are thought to be the result of a great many microscopic interactions, from which the macroscopic laws arise via the law of large numbers. Thus if the analogy between the relativity principle and the laws of thermodynamics is taken seriously, it seems we should expect that the relativity principle will eventually be given some derivation in terms of the behaviour of matter at the microscopic level. 

But if this is the status of the relativity principle, the direction of explanation favoured by Poincar\'e would surely be the appropriate one to take. Indeed, Poincar\'e himself clearly viewed the principle as a principle akin to the laws of thermodynamics. In a 1904 essay he distinguishes between the first phase of physics, where results were derived directly from the application of force laws, and the second phase of physics, where results were derived from general guiding principles. Theories belonging to this second phase are clearly what Einstein would call principle theories, and it is therefore telling that Poincar\'e includes the relativity principle, along with `Carnot's principle' (the second law of thermodynamics) in the list of the relevant principles (1904). When the relativity principle is viewed in this way, it seems correct to seek to understand it as arising out of a conjunction of more primitive hypotheses about microscopic phenomena. But Einstein's 1905 paper takes precisely the opposite approach, deriving details of the laws governing bodies from the relativity principle rather than vice versa; this suggests that whatever Einstein's personal view on the matter, the successful explanatory stategy introduced by that paper gives the relativity principle an ontological status which is not comparable to the status of the laws of of thermodynamics. 

Moreover, it is clear that in the course of the development of special relativity, the relativity principle has come to gave a significance very different from that of the laws of thermodynamics. No one would take the second law of thermodynamics to be an explanation of why microscopic bodies behave as they do, but it is common practice to see the relativity principle as an explanation for Lorentz covariance, length contraction and time dilation. I suggest that this is because the relativity principle, unlike the laws of thermodynamics, can be applied at the microscopic level. The laws of thermodynamics cannot be true of individual particles because many of the concepts to which they refer, such as temperature and entropy, are defined only with respect to the collective behaviour of particles. It is therefore appropriate to derive the laws of thermodynamics from features of microscopic behaviour, because microscopic behaviour is not itself subject to the laws of thermodynamics - it is required to produce macroscopic behaviour which obeys the laws of thermodynamics, but is not in any way explained or caused by those laws. On the other hand, the relativity principle refers only to `laws of physics' and `frames of reference,' and while we might well argue about exactly how substantive these things are, it does not seem that they become significantly more or less so as we move from the macroscopic to the microscopic level. Therefore we are able to take the principle of relativity as a necessary constraint on all laws of nature rather than merely an empirical generalisation about appearances. This is an important feature of the explanatory strategy exhibited by Einstein's 1905 paper, since, unlike Poincar\'e, he requires that laws of nature obey the principle exactly rather than conspiring to produce the appropriate appearances. But once this approach is adopted, the principle of relativity cannot be derived from the details of a microscopic theory. We can of course observe that the principle is trivially true of macroscopic phenomena simply in virtue of being true of microscopic phenomena, but we cannot derive the fact that it is true of the laws governing microscopic phenomena from the laws themselves, because we are treating the prinicple not as an accidental fact about the laws but as a universal truth about laws in general. The relativity principle has therefore come to occupy a special position in physical theory in virtue of its metatheoretical nature - we view it not merely as an empirical generalisation like the laws of thermodynamics, but as a necessary feature of all laws. 

\vspace{3mm}
\textbf{ 2.4 Explanation in the modern philosophy of special relativity}
\vspace{3mm}

Since 1905, the issue of explanation in special relativity has been made more complex by the emergence of Minkowski spacetime. There have been many questions raised about the ontological status of Minkowski spacetime and the explanatory role it should occupy. The fundamental problem is often presented as a dilemma: as Balashov and Janssen put it, `does the Minkowskian nature of space-time explain why the forces holding a rod together are Lorentz invariant or the other way around?' (2003). 

The first approach is more widely accepted. In Michael Friedman's \textit{Foundations of Space-Time Theories}, for example, the space-time manifold is understood as physically real and responsible for the spatiotemporal behaviour of matter, in particular the Lorentz covariance of the equations describing the laws of nature. But Brown objects that on this view, the role played by spacetime is mysterious: `How is its influence on these laws supposed to work? How in turn are rods and clocks supposed to know which space-time they are immersed in?' (2005) Since we know nothing about spacetime except the geometry it is apparently endowed with, and have no understanding of how it acts on and constrains the behaviour of matter, the claim that it has a certain structure offers no true explanation of the Lorentz covariance of the equations. Thus Brown prefers the alternative view: `The appropriate structure is Minkowski geometry \textit{precisely because} the laws of physics of the non-gravitational interactions are Lorentz covariant.' (2005) But this view too is subject to criticism. For if the Lorentz covariance of the equations is basic, that covariance must be a distinct hypothesis for each set of fundamental equations, and as Balashov and Janssen write, `it is, in the final analysis, an unexplained coincidence that the laws effectively governing different sorts of matter all share the property of Lorentz invariance.' (2003) In the absence of any explanation of the Lorentz covariance of the equations governing all the non-gravitational forces, it seems surprising that these independent sets of equations have exactly the same symmetry group with respect to the same frames of reference, so that they suffice to establish a unique spacetime structure. Surely there could easily have been one `spacetime' for the electromagnetic force, another for the strong nuclear force, and so on? It might be argued that these separate theories will ultimately be reducible to some `grand unified theory,' and this common source explains the common features of the equations. But the fact that distinct equations arise from the same underlying machinery does not guarantee that they will be mathematically similar in the right way, and therefore a grand unified theory would not automatically provide a satisfactory explanation of the apparent coincidence. 

I think it is significant that the the objections raised on either side of this debate closely mirror objections once levelled against Poincar\'e's theory. On the one hand, the mysteriousness of absolute space should remind us of the mysteriousness of the ether: scientists working on ether theory soon realised that that the ether `could not be a gas or a fluid, but had to be an elastic solid which had to have a high degree of rigidity to explain the high speed of light,' a troubling conclusion since `it was rather implausible that the earth and all other matter in the universe would move through a rock solid medium without in the least disturbing it.' (Jannsen and Stachel, 2008). Attempts to model the ether as a physical system seemed doomed to failure, leaving scientists with little understanding of its nature or of how it acts upon matter. The parallel between their ignorance of the ether and our current ignorance of spacetime is striking, and numerous writers have commented on the similarity: Earman writes `When relativity theory banished the ether, the space-time manifold $M$ began to function as a kind of dematerialized ether needed to support the fields,' (1989) while Brown asserts  that `the view that the space-time manifold is a substratum or bedrock, whose point elements physical fields are properties of, is just the twentieth-century version of the ether hypothesis.'(2005) On the other hand, the apparent coincidence of universal Lorentz invariance can be compared to the apparent coincidence that appears in the Lorentz-Poincar\'e theory when a number of separate fundamental laws jointly conspire to produce the required physical changes in a moving system. Poincar\'e frequently expressed concerns that Lorentz's theory, the theory he himself advocated, could preserve the relativity principle without an unsatisfactory `accumulation of hypotheses' (1904) including local time, length contraction, and the Lorentz covariance of all fundamental forces. Objecting to what he perceived as the ad hoc nature of such a collection of hypothesis, Poincar\'e wrote in 1901 that `a well formulated theory should permit proving a theory at once with all rigour. The theory of Lorentz does not permit this yet.' This is very similar to the modern objection that taking Lorentz invariance as basic requires us to postulate Lorentz invariance separately for each set of fundamental equations; once again, it would be preferable to have a means of explaning the phenomena `at once with all rigour,' rather than by a set of apparently unconnected hypotheses. 

These parallels should be taken seriously, because they suggest that two forms of explanation under consideration are both harking back to the explanatory approach of Poincar\'e. The approaches encounter the very same problems once encountered by Poincar\'e because they return to the old view that the relativity principle is somehow in need of explanation, either in terms of spacetime structure, or in terms of the Lorentz covariance of the equations. Therefore just as Einstein's 1905 paper apparently avoided the shortcomings of Poincar\'e's approach, perhaps a variant on his strategy should be applied to the modern dilemma - in particular, perhaps we should once again assume that the relativity principle itself need not be explained. On the first view we have considered here, the main function of spacetime is to guarantee the truth of the relativity principle: so if we deny that any such guarantee is required, Einstein's original 1905 derivation of the Lorentz transformations permits us to explain the Lorentz covariance of all fundamental laws without any appeal to the structure of spacetime. \footnote{It is true that Einstein's derivation requires some further empirical assumptions, such as the non-conventional part of the light-postulate, which is the claim that in at least one frame of reference the two-way speed of light is a constant. However, a similar assumption is apparently required no matter which form of explanation we adopt - for example, if we start from the structure of Minkowski spacetime, we must make the assumption that spacetime is coupled to matter and electromagnetic fields in such a way as to produce a constant two-way speed of light.} We are then free to take the structure of Minkowski as merely a codification of the behaviour of bodies as in the second approach, without having to resort to a coincidental similarity in the mathematical form of the equations. Thus the introduction of Minkowski spacetime is no reason to abandon the successful explanatory strategy introduced by Einstein's 1905 paper: spacetime structure is merely a convenient mathematical representation of the original theory, not a new element which must be separately incorporated into our explanations. 

\vspace{6mm}
\textbf{ \Large{3 Poincar\'e and Conventionalism}}
\vspace{6mm}

Arguments to the conclusion that Poincar\'e anticipated the theory of special relativity often invoke his philosophical beliefs as reasons to think he was tending towards a special relativistic interpretation of his theory. For example, Zahar claims that `Poincar\'e looks upon the effective coordinates as the only physically significant paramaters (because) he was not wedded to the classical ontology according to which absolute time, the ether frame and the Galilean coordinates are the only intelligible entities.' (2001, p. 128) But although Poincar\'e found the effective coordinates useful, he continued to see the Galilean coordinates as the true description of reality: the effective coordinates are taken to be a mistake that we make in consequence of the behaviour of objects in motion. Zahar appears to be describing not what Poincar\'e actually did, but what it seems that he logically should have done, in light of his philosophical commitments. This highlights an important feature of Poincar\'e's practice: there is often an apparent inconsistentcy in the application of philosophical principles to his scientific work. In this section I will examine some apparent contradictions between his philosophy and his science, then examine a resolution of this conflict and consider the impact of this situation on Poincar\'e's role in the development of special relativity.

\vspace{3mm}
\textbf{ 3.1 Poincar\'e's Philosophy and Poincar\'e's Science}
\vspace{3mm}

A comparison between Poincar\'e's philosophical views and his scientific theories can easily create the impression of contradiction. In particular, his theories are often grounded on ontological assumptions which he has explicitly denounced as `purely conventional' in his philosophical papers. For instance, in his 1898 paper on time, Poincar\'e makes it clear that the constancy of the one-way speed of light is a matter of convention: he writes that we begin `by supposing that light has a constant velocity, and in particular that its velocity is the same in all directions. That is a postulate without which no measurement of this velocity could be attempted.' Yet in his 1900 paper on Lorentz's theory, he describes two observers in motion relative to the ether but stationary relative to each other, and claims that: `they are not aware of their common motion, and consequently believe that the signals travel equally fast in both directions.' (1900) There is apparently a contradiction here. If the constancy of the one-way speed of light is merely a convention, the observers cannot be wrong in their belief that the signals are travelling equally fast in both directions; they are merely making a choice of convention. Darrigol (2004) suggests that the contradiction can be resolved if we assume that Poincar\'e's meaning in 1898 was that the constancy of the one-way speed of light is a convention only in the ether frame - this is certainly implied by his comment that the convention is `accepted by everyone,' (1898) because clearly, in virtue of the prevalence of the ether theory, the constancy of the speed of light in all other frames was not accepted by everyone at the time. However, I find this interpretation problematic. If the constancy of the one-way speed of light is only a convention in the ether rest frame, what meaning does it have to say that the two observers are wrong to judge that it is constant in their own frame? Presumably they are wrong with respect to the convention that it is constant in the ether rest frame – but if this is only a convention, surely there is nothing to stop them from establishing whatever convention they like in their own frame. Furthermore, a convention stipulating the speed of light in the ether rest frame only would be a very strange kind of convention, because we do not know which frame is the ether rest frame. The convention would then amount to nothing more than an agreement about how we would talk about motion with respect to different frames if, per impossible, we could distinguish the ether rest frame. Such a convention could have no consequences for our attitudes to actual empirical data. Thus it seems much more plausible to hold that Poincar\'e intended his comments to apply to observers in any frame of reference, in which case the contradiction stands. 

A similar conflict exists between Poincar\'e's philosophical views on absolute space and his use of the notion of space in his scientific work. He could not be clearer about his opinion of the notion of absolute space: `Whoever speaks of absolute space uses a word devoid of meaning.' (1897) Yet his scientific writings consistently presuppose the existence of a privileged frame of reference and suggest that only observers who are at rest in this frame of reference see phenomena as they really are. This is not an explicit contradiction, because for Poincar\'e the privileged frame of reference is merely the ether rest frame - for instance, he defines the `absolute motion of the earth' as `its motion relative to the ether instead of relative to other celestial bodies.' (Logunov 2001, p. 15). He is therefore able to talk about absolute motion without presupposing the existence of absolute space. Nonetheless, the use of the ether rest frame provides a means of avoiding the consequences of the denial of absolute space - it is essentially a way of reconciling the conventional view that there is such as thing as the true velocity and configuration of an object with the philosophical view that talk of absolute space is meaningless. This strategy therefore permits Poincar\'e to go on doing physics relative to a single frame, ignoring the interesting consequences which arise from the consideration that all frames of reference are in fact equivalent. The treatment of space in Poincar\'e's scientific work may not directly contradict his conventionalism, but it certainly seems to be in tension with the spirit of his philosophy. 

Such tensions even extend to Poincar\'e's view of the relativity principle. In his scientific papers it is consistently regarded as an empirical fact, an inductive generalisation from experiment which requires explanation in terms of more basic theories. Yet in his philosphy he claims that there are two reasons to believe in it: the first is its confirmation by experiment, but the second is that `the contrary hypothesis is repugnant to the mind,' which is presumably intended to imply that there is an a priori element to our acceptance of it. But if we have a priori reasons for believing in the principle, the demand for its explanation in terms of more fundamental theories loses much of its urgency, so if this was really Poincar\'e's view is it puzzling that he devotes so much effort to deriving it from more fundamental hypotheses - indeed, it is surprising that he never thought to do as Einstein did and take it as an axiom from which other hypotheses can be derived. 

\vspace{3mm}
\textbf{ 3.2 Resolving the contradiction}
\vspace{3mm}

The apparent contradictions between Poincar\'e's philosophical beliefs and his scientific practice can be understood as a consequence of the balance he was required to achieve between his philosophy and the practical demands of science. Clearly, Poincar\'e's conventionalism was so extreme that science could not possibly produce coherent and developed scientific theories if it were to abandon everything that he believed to be conventional. Thus Poincar\'e always makes it clear that the assertion that some rule or principle is conventional is not at all equivalent to the claim that we should cease to use that rule or principle; his approach to the circumstances in which it is appropriate to reject a convention is much more nuanced. In his 1904 esssay, `The Future of Mathematical Physics,' he discusses this issue, observing that although experimental results can never contradict a convention, the convention will nevertheless be threatened if in order to retain it we are forced to add ad hoc hypotheses such that it ceases to be predictively useful - for instance, the attempt to explain the apparent violation of conservation of energy in the case of radiation from radium by the hypothesis that unobservable quantities of energy are constantly travelling through space in all directions, and some of this energy is converted into an observable form inside radium. Poincar\'e is of the opinion that in such circumstances the relevant convention should be abandoned, but we should `abandon the conventions only after having made a loyal effort to save them.' (1904) Indeed, in papers such as his 1900 contribution to the Lorentz Festschrifte, he makes a valiant attempt to preserve the principle of action-reaction in the face of its apparent violation by electrodynamic phenomena. 

Poincar\'e often refers to the network of conventions used in science as a framework; so for instance he claims that 'space is another framework which we impose on the world.' The fact that all his scientific work presupposes the existence and reality of space therefore reflects a deliberate choice to work within the traditional framework of science. Indeed, his commitment to retaining conventions whenever possible makes it clear that he does not see it as the place of the scientist to question this framework; as Stein observes, `the basic mathematical presuppositions of physics were seen by Poincar\'e as defining a framework within which \textit{it is the task of the theoretical physicist to fit all phenomena.}' (unpublished). With this in mind, we can understand why Poincar\'e insisted on presupposing long-established scientific conventions in his own scientific work, for example, by producing explanations which conform to the traditional explanatory strategy of explaining macroscopic phenomena by appeal to the motions of microscopic particles in absolute space.  He recognised that this explanatory framework is a freely chosen convention, but he believed that the role of science is to construct theories within this framework, not to investigate the nature of the framework itself. 

Einstein, on the other hand, is much more willing to discard conventions - perhaps the clearest example is his willingness to adopt a new synchrony convention which violated traditional ideas about the nature of time. Poincar\'e would theoretically have agreed with Einstein that simultaneity is determined by a synchrony convention, but unlike Einstein he always retained the traditional conventions in his scientific work. However, we should be aware that Einstein certainly did not get rid of everything that he believed conventional. For example, he retains the fiction that distances and coordinates within a single frame have objective, determinate values, even though he acknowledges that positions can only be defined `by the employment of rigid standards of measurement,' (1905) which suggests that he had an awareness of the conventional nature of such standards. Thus both Poincar\'e and Einstein recognised that science can progress only with the support of a network of descriptive conventions; the difference is merely that Einstein was willing to be more flexible about which conventions are retained. 

\vspace{3mm}
\textbf{ 3.3 The Convenience Thesis}
\vspace{3mm}

One feature of Poincar\'e's conventionalism that makes it particularly nuanced and interesting is his insistence that experience guides us in choosing conventions: for instance, after arguing that the geometry of space is a matter of convention, he writes that `experiment ... tells us not which is the truest, but which is the most convenient geometry.' (1902, p. 71-72) Similarly, after arguing that the laws of acceleration and composition of forces are conventions, he writes that `they would seem arbitrary if we forgot the experiences which guided the founders of science to their adoption and which are, although imperfect, sufficient to justify them.' (1902, p. 110) As Ben-Menahem puts it, Poincar\'e `critiques both an oversimplied conception of fact and an equally oversimplified conception of convention,' and this gives his conventionalism much greater plausibility than versions upon which conventions are taken to be arbitrary. 

However, I think there are two related theses in this vicinity, which neither Poincar\'e nor later commentators have adequately distinguished. The first is the claim that not all conventions are equally good: as Ben-Menahem puts it, `the choice of a coordinate system or measurement unit is intricately linked to obective features of the situation.' Thus we can accept that certain conventions cannot be judged true or false, but still argue that it is possible to make rational choices between conventions for practical reasons. The second is what I will call the convenience thesis: the actual process by which conventions come to be selected is such that the conventions we ultimately choose are the most convenient, so in the sciences we somehow end up selecting the conventions which allow us to express the laws of nature in the simplest possible way. The two theses are frequently treated together, as if the very fact that we can make a reasoned choice between conventions implies that the historical process by which conventions are determined will always select the most appropriate conventions. But the convenience thesis by no means follows immediately from the claim that not all conventions are equal, and indeed, I suggest that we have little reason to think it is true.  

The convenience thesis appears to neglect the fact that our conventions are not easily altered, and certainly do not change in step with our understanding of the laws of nature. The conventions regarding the measurement of distance and time used by physicsts like Lorentz and Poincar\'e were essentially the same conventions that had been used for thousands of years - yet there seems no reason to suppose that the conventions which permitted the simplest expression of the laws of nature as our ancestors understood them would still permit the simplest expression of the laws of nature as understood in 1900. The early history of special relativity provides a clear illustration of this problem: in retrospect, we can see that discarding certain conventions about synchrony and simultaneity allows us to attain greater simplicity both in the form of our fundamental equations and also in the structure of the explanations we are able to offer, but there is no way this could have been taken into account in the original formation of our conventions regarding space and time, because it is a consequence of features of electrodynamical theory such as the invariance of the speed of light, and this theory was completely unknown at the time the conventions were formed. Thus one conclusion to be drawn from this episode is that even if conventions are somehow selected in such a way as to be most convenient for the purposes of science at the time of their creation, it does not follow that the same conventions will still be the most convenient after science has had a chance to develop. 

Moreover, Poincar\'e's trust in the simplicity thesis was certainly one of the major factors that prevented him from seeing the possibility of using the relativity principle as explanans rather than explanandum. The convenience thesis is in itself no more than a historical claim, but it naturally leads to a number of normative claims about how scientists should behave with respect to matters which they believe to be conventional. If it is accepted that we naturally come to adopt the conventions which permit the simplest expression of the laws of nature, this provides a powerful motivation to retain the conventions that we currently have, and consequently, Poincar\'e held that it is the role of science to work within the established framework of conventions rather than question that framework. In light of this view, it is entirely comprehensible that Einstein's approach did not occur to Poincar\'e. Einstein's success arose from the insight that by abandoning certain conventions we may achieve a great simplification in the structure of our explanations; but Poincar\'e believed such that such conventions had come to be selected precisely because they enabled us to give the simplest formulation of the laws of nature, so it is not surprising that he did not anticipate that rejecting a convention could lead to further simplifications.

\vspace{6mm}
\textbf{ \Large{Conclusions}}
\vspace{6mm}

Poincar\'e's contributions to the field of special relativity were undoubtedly invaluable, but nonetheless those contributions do not constitute an independent discovery of the theory. His conceptual grasp of certain elements, particularly the Lorentz transformations, was very different to Einstein's - and this does not constitute merely a small difference in interpretation, but a substantive disagreement about the possible applications of the transformations. 

I have suggested that the root of this conceptual difference lies in a disagreement over the appropriate direction of explanation; for Poincar\'e, the relativity principle is to be explained by invoking microscopic phenomena, whereas Einstein showed that the relativity principle can be taken as basic and used to explain the form of the laws governing microscopic phenomena. This has important consequences for issues concerning explanation in special relativity – a return to the approach of 1905 may help to resolve issues concerning the role of Minkowski spacetime in relativistic explanations. 

Finally, I have argued that Poincar\'e's choice to restrict his scientific explanations to traditional forms was related to certain conventionalist doctrines that he held, particularly the convenience thesis. This thesis led Poincar\'e to maintain that science should not question the conventional framework within which it is mandated to work, and as a consequence, he was unwilling to give up the conventional framework of explanation. Therefore an explanatory approach like Einstein's was unavailable to him and he never appreciated the interesting possibilities that arise from a willingness to explain in non-traditional ways.

\vspace{6mm}
\textbf{ \Large{Acknowledgements}}
\vspace{6mm}

I would like to thank Professor Harvey Brown for his support and valuable input throughout the production of this paper.

\vspace{6mm}
\textbf{ \Large{References:}} 
\vspace{6mm}

Achinstein, P. (1983). \textit{The Nature of Explanation}. Oxford: Oxford University Press.

\vspace{2mm}

Aristotle. \textit{Physics}. English translation in Hardie, R. and Gaye, R. (1930). \textit{Physica}. The Clarendon Press, Oxford. 

\vspace{2mm} 

Barbour, J. (2001). \textit{The Discovery of Dynamics}. Oxford: Oxford University Press. 

\vspace{2mm}

Ben-Menahem, Y. (2006). \textit{Conventionalism}. Cambridge: Cambridge University Press. 

\vspace{2mm}

Brown, H. (2005). \textit{Physical Relativity}. Oxford: Oxford University Press. 

\vspace{2mm}

Brush, S. (1999). \textit{Why was Relativity Accepted?} Physics in Perspective vol 1, p. 184–214.
\vspace{2mm}

Campbell, N. (1920). \textit{Physics: The Elements}. Cambridge: Cambridge University Press. 

\vspace{2mm}

Darrigol, O. (2004). \textit{The Mystery of the Einstein - Poincar\'e Connection}. Isis, vol 95 (4), p. 614–626.

\vspace{2mm}

Earman, J. (1989) \textit{World Enough and Space-Time: Absolute versus Relational Theories of Space and Time}. Cambridge, MA: MIT Press.

\vspace{2mm}

Einstein, A. (1905).\textit{ On a Heuristic Viewpoint Concerning the Production and Transformation of Light}. Annalen der Physik, vol 17 p. 132–148.

\vspace{2mm}

Einstein, A. (1905). \textit{On the Electrodynamics of Moving Bodies}. Annalen der Physik, vol 17 p. 891-921.

\vspace{2mm}

Einstein, A. (1949). \textit{Autobiographical Notes}. In \textit{Albert Einstein, Philosopher--Scientist}. (1959) Paul Arthur Schilpp, Ed. New York: Harper and Bros.

\vspace{2mm}

Einstein, A. and Infeld, L. (1961) \textit{The Evolution of Physics}. New York: Simon and Schuster.

\vspace{2mm}

Friedman, M. (1986). \textit{On The Foundations of Spacetime Theories}. Princeton: Princteon University Press.

\vspace{2mm}

Goldberg, S. (1967). \textit{Henri Poincaré and Einstein's Theory of Relativity}. American Journal of Physics, vol 35, p. 934-44

\vspace{2mm}

Hempel, C. (1966). \textit{Philosophy of Natural Science}. Prentice-Hall, Englewood Cliffs.

\vspace{2mm}

Hirosige, T. (1976). \textit{The Ether Problem, the Mechanistic Worldview, and the Origins of the Theory of Relativity}. Historical Studies in the Physical Sciences, vol 7, p. 3-82.

\vspace{2mm}

Janssen, M. (1995). \textit{A Comparison between Lorentz's Ether Theory and Special Relativity in the Light of the Experiments of Trouton and Noble}. Dissertation. University of Pittsburgh, 1995.

\vspace{2mm}

Janssen, M. and Stachel, J. (2008). \textit{The Optics and Electrodynamics of Moving Bodies}. Preprint.
 http://www.mpiwg-berlin.mpg.de/Preprints/P265.PDF

\vspace{2mm}

Katzir, S. (2005). \textit{Poincar\'e's Relativistic Physics: Its Origins and Nature}. Physics in Perspective, vol 7, p. 268–292

\vspace{2mm}

Lorentz, H. (1899). \textit{Simplified Theory of Electrical and Optical Phenomena in Moving Systems}. Proc. Acad. Science Amsterdam 1, 427–442.

\vspace{2mm}

Lorentz, H.A (1916). \textit{The Theory of Electrons}. Leipzig and Berlin.

\vspace{2mm}

Miller, A.  (1973). \textit{A Study of Henri Poincaré's 'Sur la Dynamique de l'Électron}. 
Archive for History of the Exact Sciences, vol 10, p. 320

\vspace{2mm}

Norton, J. (1986). \textit{The Quest for the One Way Velocity of Light}. British Journal for the Philosophy of Science, vol 37: 118–120.

\vspace{2mm}

Pais, A. (1982). \textit{Subtle Is the Lord: The Science and the Life of Albert Einstein}. New York: Oxford University Press. 

\vspace{2mm}

Planck, M. (1906). \textit{The Principle of Relativity and the Fundamental Equations of Mechanics}. Verhandlungen Deutsche Physikalische Gesellschaft. 8, p. 136–141.

\vspace{2mm}

Poincar\'e, H. (1898). \textit{The Measure of Time}. English translation in \textit{The Foundations of Science (The Value of Science)} (1913) New York: Science Press. p. 222-234.

\vspace{2mm}

Poincar\'e, H. (1900). \textit{The Theory of Lorentz and the Principle of Reaction} Archives nèerlandaises des Sciences exactes et naturelles, series 2, volume 5, pp 252-278.

\vspace{2mm}

Poincar\'e, H. (1901). \textit{L'electricite et optique: La lumiere et les theories electrodynamiques}. Paris: Gauthier-Villars.

\vspace{2mm}

Poincar\'e, H. (1902). \textit{Science and Method}. English translation in \textit{Science and Method}. (1952) London: Dover Publications Inc. 

\vspace{2mm}

Poincar\'e, H. (1904). \textit{L'etat et l'avenir de la Physique mathematique}. Bulletin des Sciences Mathematiques, vol 28 p.302-324.

\vspace{2mm}

Poincar\'e, H. (1906). \textit{Sur la dynamique de l'electron}. Rendiconti del Circolo matematico di Palermo, vol 21, p. 129-175. English translation in Logunov, A. (2001). \textit{On the Articles by Henri Poincar\'e `On The Dynamics of the Electron'}. Dubna: Joint Institute for Nuclear Research. 

\vspace{2mm}

Stein, H. {Physics and Philosophy Meet: The Strange Case of Poincar\'e} University of Chicago, unpublished.
 http://strangebeautiful.com/other-texts/stein-strange-case-poincare.pdf

van Fraassen, B. C. (1980). \textit{The Scientific Image}. Oxford: Clarendon Press.

\vspace{2mm}

Zahar, E. (2001). \textit{Poincar\'e's Philosophy}. Chicago: Carus Publishing Company. 

\vspace{2mm}

Zahar, E. (1989). \textit{Einstein's Revolution}. Chicago: Open Court Company.

\end{document}